\documentclass[prl,aps,twocolumn,showpacs,superscriptaddress]{revtex4}
\usepackage{amsbsy}
\usepackage{bm}
\usepackage{graphicx}

\newcommand{\nc}{\newcommand}
\nc{\noi}{\noindent}     
\nc{\eq}[1]{\mbox{Eq.~(\ref{#1})}}
\nc{\ba}{\begin{array}}
\nc{\ea}{\end{array}}
\nc{\bea}{\begin{eqnarray}}
\nc{\eea}{\end{eqnarray}}
\nc{\fig}[1]{\mbox{Fig.~\ref{#1}}}

\begin{document}

\title{Spontaneous Emission Near Superconducting Bodies}

\author{Bo-Sture K. Skagerstam}\email{bo-sture.skagerstam@ntnu.no}
\author{Per Kristian Rekdal}\email{pkrekdal@gmail.com}
\affiliation{Complex Systems and Soft Materials Research Group, Department of Physics, 
             The Norwegian University of Science and Technology, N-7491 Trondheim, Norway}

\date{\today}

\begin{abstract}

     In the present paper we study the spontaneous photon emission due to a magnetic spin-flip transition
     of a two-level atom in the vicinity of a dielectric body like a normal conducting metal or
     a superconductor. For temperatures below the transition temperature $T_c$ of a superconductor,
     the corresponding spin-flip lifetime is boosted by several orders of magnitude as
     compared to the case of a normal conducting body. Numerical results of an exact formulation are also
     compared to a previously derived approximative analytical
     expression for the spin-flip lifetime and we find an excellent agreement.
     We present results on how the spin-flip lifetime depends on the temperature $T$ of a superconducting body
     as well as its thickness $H$. Finally, we study how non-magnetic impurities as well as possible Eliashberg
     strong-coupling effects influence the spin-flip rate.
     It is found that non-magnetic impurities as well as strong-coupling effects have no dramatic impact
     on the spin-flip lifetime.

\end{abstract}


\maketitle

     It is well-known that the rate of spontaneous emission of atoms will be modified due to the presence
     of a dielectric body \cite{purcell_46}. In current investigations of atom microtraps this issue is of  
     fundamental importance since such decay processes have a direct bearing on the stability of e.g. atom chips.

     In magnetic microtrap experiments, cold atoms are trapped due to the presence of  magnetic field gradients created e.g. by 
     current carrying wires \cite{folman_02}. 
     Such microscopic traps provide a powerful tool for the control and manipulation of
     cold neutral atoms over micrometer distances \cite{henkel_06}.
     Unfortunately, this proximity of the cold atoms to a dielectric body introduces additional decay channels. 
     Most importantly, Johnson-noise currents in the material give rise to electromagnetic field fluctuations. 
     For dielectric bodies at room temperature made of normal conducting metals,
     these fluctuations may be strong enough to deplete the quantum state of the atom and, hence,
     expel the atom from the magnetic microtrap \cite{jones_03}.
     Reducing this disturbance from the surface is therefore strongly desired. 
     In order to achieve this, the use of superconducting dielectric bodies instead of
     normal conducting metals has been proposed \cite{scheel_05}. Some experimental work in this context
     has been done as well, e.g. by Nirrengarten {\it et al.} \cite{haroche_06}, where cold atoms
     were trapped near a superconducting surface.

     In the present article we will  consider the spin-flip rate when the electrodynamic properties
     of the superconducting body are described in terms of either a simple two-fluid model or
     in terms of the detailed microscopic Mattis-Bardeen \cite{mattis_58} and
     Abrikosov-Gor'kov-Khalatnikov \cite{gorkov58} theory of weak-coupling BCS superconductors.
     In addition, we will also study how non-magnetic impurities, 
     as well as strong coupling effects according to the low-frequency limit of
     the Eliashberg theory \cite{eliashberg_60}, will affect the spontaneous emission rate.

     Following Ref.\cite{rekdal_04} we consider an atom in an initial state $|i \rangle$ and trapped at position ${\bf r}_A= (0,0,z)$
     in vacuum near a dielectric body. The rate $\Gamma_B$ of spontaneous and thermally stimulated magnetic spin-flip
     transition into a final state $|f\rangle$ is then 
\bea \label{gamma_B_generel}
     \lefteqn{\Gamma_B = \, \mu_0 \,  \frac{2 \, (\mu_B g_S)^2}{\hbar} \; \sum_{j,k} ~  
     S_j \, S_k^{\, *} }
     \nonumber 
     \\ 
     && \; \times \;
     \mbox{Im}  \, [  \; \nabla \times \nabla \times
     \bm{G}({\bf r}_A  ,  {\bf r}_A  ,  \omega ) \; ]_{jk}  \;  ( {\overline n} + 1 ) \, ,
\eea
     where we have introduced the dimensionless components $S_j \equiv  \langle f | \hat{S}_j/\hbar | i \rangle$
     of the  electron spin operators $\hat{S}_j$ with $j=x,y,z$. 
     Here $g_S \approx 2$ is the gyromagnetic factor of the electron,
     and $\bm{G}({\bf r},{\bf r}',\omega )$ is the dyadic Green tensor of Maxwell's theory. 
     \eq{gamma_B_generel} follows from a consistent quantum-mechanical treatment of electromagnetic 
     radiation in the presence of an absorbing body \cite{dung_00,henry_96}. In this theory a 
     local response is assumed, i.e. the characteristic skin depth should be larger than the mean free path
     of the electric charge carriers of the absorbing body.
     Thermal excitations of the electromagnetic field modes are accounted for by the factor 
     $( {\overline n} + 1 )$, where ${\overline n} = 1 / ( e^{\hbar \omega/k_{\text{B}} T}- 1 )$ and
     $\omega \equiv 2 \pi \, \nu$ is the angular frequency of the spin-flip transition. 
     Here $T$ is the temperature of the dielectric body, which is assumed to be in thermal equilibrium
     with its surroundings. 
     The dyadic Green tensor is the unique solution to the Helmholtz equation 
\bea   \label{G_Helm}
   \nabla\times\nabla\times \bm{G}({\bf r},{\bf r}',\omega) - k^2
   \epsilon({\bf r},\omega) \bm{G}({\bf r},{\bf r}',\omega) = \delta( {\bf r} - {\bf r}' ) \bm{1} \, ,\nonumber\\
\eea
     with appropriate boundary conditions. Here $k=\omega/c$ is the wavenumber in vacuum, $c$ is the speed of light 
     and $\bm{1}$ the unit dyad. The tensor $\bm{G}({\bf r},{\bf r}',\omega)$ contains all 
     relevant information about the geometry of the material and, through the relative electric permittivity
     $\epsilon({\bf r},\omega)$, about its dielectric properties.
     The fluctuation-dissipation theorem is build into this theory \cite{dung_00,henry_96}.


    The decay rate $ \Gamma^{\, 0}_{ B } $ of a magnetic spin-flip transition for an atom in
    free-space is well-known (see e.g. Refs.\cite{dung_00}). This free-space decay rate is
    $\Gamma^{\, 0}_B   = \Gamma_B   S^{\, 2}$, where $\Gamma_B  =  \mu_0  \, ( \mu_B g_S )^2 \, k^3 / ( 3 \pi  \hbar )$
    and where we have introduced the dimensionless spin factor $S^{\, 2} \equiv S_x^{\, 2} + S_y^{\, 2} + S_z^{\, 2}$. 
    The free-space lifetime corresponding to this magnetic spin-flip rate is $\tau^{\, 0}_{ B } \equiv 1/\Gamma^{\, 0}_{ B }$.
    In the present paper we only consider $^{87}$Rb atoms that are initially pumped into the
    $|5 S_{1/2},F=2,m_F=2\rangle \equiv |2,2\rangle$ state, and assuming the rate-limiting transition
    $|2,2 \rangle \rightarrow |2,1 \rangle$ in correspondence to recent experiments \cite{vuletic_04,harber_03,hinds_03,haroche_06}.
    The spin factor is $S^2=1/8$ (c.f. Ref.\cite{rekdal_04}) and the frequency is $\nu = 560$ kHz.
    The numerical value of the free-space lifetime then is $\tau^0_B = 1.14 \times 10^{25}$ s.

    In the following we will consider a geometry where an atom is trapped 
    at a distance $z$ away from a dielectric slab with thickness $H$.
    Vacuum is on both sides of the slab, i.e. $\epsilon({\bf r},\omega) = 1$
    for any position ${\bf r}$ outside the body. The slab can be e.g. a superconductor or
    a normal conducting metal, described by a dielectric function $\epsilon(\omega)$. 
    The total transition rate for magnetic spontaneous emission
\bea  \label{total_B}
  \Gamma_{ B } =  ( \Gamma^{\, 0}_{B} +  \Gamma^{\, \rm{slab}}_{B} ) \, ( {\overline n} + 1 ) \, ,
\eea
    can then be decomposed into a free part and a part purely due to the presence of the slab.
    The latter contribution for an arbitrary spin orientation is then given by
\bea
    \Gamma^{\, \rm{slab}}_{B}  &=&  \label{gamma_B_gen_ja}
    2 \, \Gamma^{\, 0}_{B}   \, 
    \bigg ( ~  (  S_x^{\, 2} \, + \, S_y^{\, 2}  ) \, I_{\|}   \, + \, S_z^{\, 2} \, I_{\perp} ~ \bigg )  ~ , 
\eea
    \noi
    with the atom-spin orientation dependent integrals 
\bea \nonumber
  I_{\|}   &=& 
                \frac{3}{16 k z} \, 
                \textrm{Re}  
                \bigg \{ 
        \int_{0}^{2 k z}  d x \, e^{ i  x }  
        \bigg [  {\cal C}_{N}(x)  -  ( \frac{x}{2 k z}  )^2  {\cal C}_{M}(x)  \bigg ]  
      \\ \label{I_para}
      && + \,  \int_{0}^{\infty}  d x \,  e^{ - x }  \frac{1}{i}
        \bigg [  {\cal C}_{N}(ix)   +  ( \frac{x}{2 k z} )^2  {\cal C}_{M}(ix)  \bigg ]  \bigg \} \, ,
     \\ \nonumber
     \\ \nonumber
  I_{\perp}   &=& 
                \frac{3}{8 k z} \, 
                \textrm{Re} 
                \bigg \{ \,
        \int_{0}^{2 k z}  d x  \, e^{ i x } \, 
        \bigg [ 1 - ( \frac{x}{2 k z} )^2  \bigg ] \, {\cal C}_{M}(x) 
      \\ \label{I_perp}
      && ~~~~~  + ~  \int_{0}^{\infty}  d x \,  e^{ - x } \, \frac{1}{i} \, \bigg [  1 + ( \frac{x}{2 k z} )^2   \bigg ] \, 
                     {\cal C}_{M}(ix)  \bigg \}  ,
\eea
   \noi
   where the scattering coefficients are given by \cite{li_94}
\bea 
     {{\cal C}}_{N}(x)   =     r_p(x) ~ \frac{1 - e^{ix \, H/z} }
                                             {1 \, - \, r_p^2(x) \, e^{ix \, H/z} }   ~  ,  
\\ \nonumber
\\ 
  {{\cal C}}_{M}(x)   =    r_s(x) ~ \frac{1 - e^{ix \, H/z} }
                                              {1 \, - \, r_s^2(x)  \, e^{ix \, H/z} }  ~ .  
\eea
   \noi
   The electromagnetic field polarization dependent Fresnel coefficients are
\bea 
       r_p(x)  &=&    \frac{\epsilon(\omega) \, x  \, - \,  \sqrt{\, ( 2 k z )^2 ( \, \epsilon(\omega) - 1 \, ) \, + \, x^2} }
                           {\epsilon(\omega) \, x \, + \, \sqrt{\, ( 2 k z )^2 ( \, \epsilon(\omega) - 1 \, ) \, + \, x^2} } ~ ,
\\ \nonumber
\\  \label{r_s}
  r_s(x)  &=&    \frac{ x \, - \, \sqrt{\, ( 2 k z )^2 ( \, \epsilon(\omega) - 1 \, ) \, + \, x^2} }
                                       { x \, + \, \sqrt{\, ( 2 k z )^2 ( \, \epsilon(\omega) - 1 \, ) \, + \, x^2} } ~ . 
\eea
    \noi
    For the special case $H=\infty$, the integrals in Eqs.(\ref{I_para}) and (\ref{I_perp}) are simply a convenient
    re-writing of Eqs.(8)-(12) in Ref.\cite{rekdal_07}. Note that $I_{\perp} \approx 2 \, I_{\|}$ provided $k z \ll 1$.
    Throughout this article, we use the same spin-orientation as in Refs.\cite{scheel_05,rekdal_06}, i.e. $S_y^2 = S_z^2$ and $S_x=0$.

\begin{figure}[ht]

\begin{picture}(0,0)(122,665) 

\includegraphics{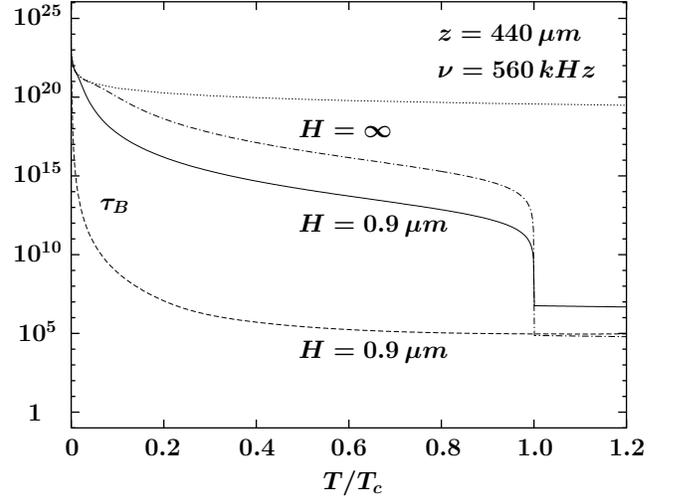}

\end{picture}

\vspace{6.5cm}

\caption{$\tau_B$ of a trapped atom near a superconducting film as a function of the temperature $T/T_c$.
         The solid as well as the dashed-dotted line correspond to the
         two-fluid mode and the Gorter-Casimir temperature dependence. We use $\lambda_L(0) = 35 \,$nm and
         $\delta(T_c) \approx 150 \, \mu$m \cite{pronin_98}, corresponding to niobium. The critical
         temperature is $T_c=8.31$ K \cite{pronin_98}. For $T/T_c \ge 1$ we put $\sigma_2(T) \simeq 0$
         but $\sigma_1 (T)=2/\omega\mu_0 \delta(T_c)^2$.
         The dashed line corresponds to a film made of gold described by the dielectric function given by
         \eq{eps_iver}.
         The upper most graph (dotted line) shows to the lifetime $\tau_B^0/( {\bar n} + 1 )$, i.e. the 
         free-space lifetime with $\tau_{B}^0 =1.114\times 10^{25}$.
}
\label{tau_sfa_t}
\end{figure}

    As the total current density responds linearly and locally to the electric field,
    the dielectric function can be written
\bea \label{eps_j_2}
  \epsilon(\omega) = 1 - \frac{\sigma_2(T)}{\epsilon_0 \, \omega} + i \, \frac{\sigma_1(T)}{\epsilon_0 \, \omega} \, .
\eea
     Here $\sigma(T) \equiv \sigma_1(T) + i \sigma_2(T)$ is the complex optical conductivity.
     We may now parameterize this complex conductivity in terms of the London penetration length 
     $\lambda_L(T) \equiv \sqrt{1/\omega \mu_0 \sigma_2(T)}$ and the skin depth
     $\delta (T) \equiv \sqrt{2/\omega \mu_0 \sigma_1(T)}$.
     In this case, the dielectric function is $\epsilon(\omega) = 1 - 1/k^2 \lambda^2_L(T) + i \, 2/k^2\delta^2(T)$.
     If, in addition, we consider a non-zero and sufficiently small frequency
     in the range $0 < \omega \ll \omega_g \equiv 2 \Delta(0)/\hbar$,
     where $\Delta(0)$ is the energy gap of the superconductor at zero temperature,
     the current density may be described in terms of a two-fluid model \cite{london_34_40}.
     The London penetration length is $\lambda_L(T) = \lambda_L(0)/\sqrt{ n_s(T)/n_0 }$
     and the skin depth is $\delta(T) = \delta(T_c)/\sqrt{n_n(T)/n_0}$.
     Here the electron density in the superconducting and normal state are $n_s(T)$ and $n_n(T)$, respectively,
     such that $n_s(T) + n_n(T) = n_0$ and $n_s(0) = n_n(T \geq T_c) = n_0$ \cite{london_34_40}.
     A convenient summary of the two-fluid model is expressed by the relations
\bea \label{tfm_sigma1_2}
  \sigma_1(T) = \sigma_n \sqrt{\frac{n_n(T)}{n_0}} ~~ , ~~  \, \sigma_2(T) = \sigma_L \sqrt{\frac{n_s(T)}{n_0}} \, ,
\eea
     where $\sigma_n \equiv \sigma_1(T_c)$ and $\sigma_L \equiv 1/\omega\mu_0\lambda^2_L(0)$. 
     Considering, in particular, the Gorter-Casimir temperature dependence \cite{gorter_34}
     for the current densities, the electron density in the normal state is $n_n(T)/n_0 = (T/T_c)^4$.
     For niobium we use $\delta(T_c) = \sqrt{2/\omega \mu_0 \sigma_n} \approx 150 \, \mu$m as
     $\sigma_n \approx 2 \times 10^7 (\Omega \textrm{m})^{-1}$ and 
     $\lambda_L(0) = 35 \,$nm according to Refs.\cite{pronin_98}. In passing, we remark that
     the value of $\sigma_n$ as obtained in Ref.\cite{casalbuoni_05} is two orders of magnitude
     larger than the corresponding value inferred from the data presented in Refs.\cite{pronin_98}. 

     The lifetime $\tau_B \equiv 1/\Gamma_B$ for spontaneous emission as a function of $T$ is shown in \fig{tau_sfa_t}
     for $H=0.9 \, \mu$m (solid line). We confirm the observation in Ref.\cite{rekdal_06} that for temperatures
     below $T_c$ and for $H=\infty$ (dash-dotted line), the spin-flip lifetime  is boosted
     by several orders of magnitude. 
     In Ref.\cite{rekdal_06}, the spin-flip lifetime was, however, calculated by making use of the
     approximative and analytical expression
\bea \label{skagerstam06}
     \frac{\tau^B_0}{\tau^B } =  ( {\bar n} + 1 )\left(  1 + (\frac{3}{4})^3 \sqrt{\epsilon_0 \omega} \,
                                 \frac{\sigma_1(T)}{\sigma_2^{3/2}(T)} \, \frac{1}{(kz)^{4}} \right) \, ,
\eea
    valid provided $\lambda_L(T) \ll \delta (T)$ and $\lambda_L(T) \ll z \ll \lambda$.
    Comparing this analytical expression with the numerical results as presented in \fig{tau_sfa_t},
    based on the exact equations Eqs.(\ref{total_B})-(\ref{r_s}), we find an excellent agreement. This observation remains true
    when $\sigma_1 (T)$ and $\sigma_2 (T)$ are obtained from more detailed and microscopic considerations to be discussed below.
    For temperatures $T/T_c > 1$ we can neglect the  $\sigma_2 (T)$ dependence and, for $\delta (T) \ll z$,
    we confirm the result of Ref.\cite{scheel_05}, i.e.
\bea \label{knight05}
     \frac{\tau^B_0}{\tau^B } =  ( {\bar n} + 1 )\left(  1 + (\frac{3}{4})^3 \sqrt{\frac{2\epsilon_0 \omega}{\sigma_1(T)}}
                                 \, \frac{1}{(kz)^{4}} \right) \, .
\eea
    For $T \simeq T_c$ we have to resort to numerical investigations.

    In contrast to the traditional Drude model, more realistic descriptions of a normal conducting metal in
    terms of a permittivity include a significant real contribution to the dielectric function
    in addition to an imaginary part.
    One such description is discussed in Ref.\cite{brevik_05}, where
\bea \label{eps_iver}
  \epsilon(\omega,T) = 1 - \frac{\omega_p^2}{\omega^2 + \nu(T)^2} + i \, \frac{\nu(T) \, \omega_p^2}{\omega \, ( \, \omega^2 + \nu(T)^2 \, ) } \, ,
\eea
    and $\hbar\nu(T) = 0.0847 (T/\theta)^5\int_0^{\theta/T} dx \, x^5  e^x /(e^x - 1)^2$ eV using a Bloch-Gr\"{u}neisen
    approximation. Here $\theta=175$ K for gold. The plasma frequency is $\hbar \omega_p = 9$ eV. 
    For temperatures $T\simeq 0.25 T_c$, we observe that Eq.(\ref{eps_iver}) leads to $\sigma_1(T) \simeq \sigma_2(T)$,
    and that for lower temperatures $\sigma_2(T)$ will be the dominant contribution to the conductivity.
    For temperatures $T/T_c \gtrsim 1$, in the use of Eq.(\ref{eps_iver}) we can set $\sigma_2(T) \simeq 0$ when
    calculating the lifetime. For a bulk material of gold this leads to almost two orders of magnitude longer lifetime
    as compared to niobium since for gold $\delta(T_c) \approx 1 \, \mu$m, using the parameters corresponding
    to Fig.\ref{tau_sfa_t}. This finding is in accordance with Eq.(\ref{knight05}).
    As seen from Fig.\ref{tau_sfa_t}, for a thin film and for $T/T_c \geq 1$ we find the opposite and remarkable result,
    i.e. a decrease in conductivity can lead to a larger lifetime.


    A much more detailed and often used description of the electrodynamic properties of superconductors than
    the simple two-fluid model was developed by Mattis-Bardeen \cite{mattis_58}, and independently by
    Abrikosov-Gor'kov-Khalatnikov \cite{gorkov58},  based on the weak-coupling BCS theory of superconductors. 
    In the clean limit, i.e. $l \gg \xi_0$, where $l$ is the electron mean free path and $\xi_0$ is the coherence
    length of a pure material, the complex conductivity, normalized to $\sigma_n \equiv \sigma_1(T_c)$,
    can be expressed in the form \cite{klein_94}
\bea  \nonumber
  \frac{\sigma(T)}{\sigma_n} &=& \int_{\Delta(T) - \hbar \omega}^{\infty} \frac{dx}{\hbar \omega} \, 
        \tanh \bigg ( \frac{x+\hbar \omega}{2 k_B T} \bigg ) \, g(x)
  \\   \label{sigma_klein}
  &-&  \int_{\Delta(T)}^{\infty} \frac{dx}{\hbar \omega} \, \tanh \bigg ( \frac{x}{2 k_B T} \bigg ) \, g(x) \, ,  ~~~~~~~
\eea
   \noi
   where $g(x) = ( x^2 + \Delta^2(T) + \hbar \omega \, x )/u_1 \, u_2$
   and $u_1 = \sqrt{x^2 - \Delta^2(T)}$, $u_2 = \sqrt{(x + \hbar \omega)^2 - \Delta^2(T)}$. 
   Here, the well-known BCS temperature dependence for the superconducting energy gap $\Delta(T)$ is given by \cite{fetter_71}
\bea  \nonumber
   &&\ln{  \left [ ~ \bigg( \, \hbar \omega_D + \sqrt{(\hbar \omega_D)^2  + \Delta^2(0)} \, \bigg) / \Delta(0) ~ \right ]  }
   \\  \label{trans_Debye}
   && =
   \int_0^{\hbar \omega_D}  \frac{d x}{\sqrt{x^2 + \Delta^2(T)}} \, \tanh \left[ \, \frac{\sqrt{x^2 + \Delta^2(T)}}{2 \, k_B T} \, \right ] \, , ~~~~
\eea
    where $\omega_D$ is the Debye frequency and $\Delta(0) = 3.53 \, k_B T_c/2$.
    For niobium, the Debye frequency is  $\hbar \omega_D = 25$ meV.
    According to a theorem of Anderson \cite{anderson_59,abrikosov_58}, the presence of non-magnetic impurities,
    which we only consider in the present paper, will not modify the superconducting energy gap as given by \eq{trans_Debye}.
    The complex conductivity will, however, in general be modified due to the presence of such impurities.

    In the dirty limit where $l \ll \xi_0$, the complex conductivity has been examined within the framework
    of the microscopic BCS theory (see e.g. Ref.\cite{chang_89}). In this case, the complex conductivity,
    now normalized to $\sigma_L$, can conveniently be written in the form
\bea  \nonumber
  \frac{\sigma(T)}{\sigma_L}  &=& 
   \int_{\Delta(T) - \hbar \omega}^{\infty} \frac{dx}{2}  ~ \tanh \bigg ( \frac{x+\hbar \omega}{2 k_B T} \bigg ) 
   \\  \nonumber
   &&  \times  \bigg (  \frac{g(x) + 1}{u_2 - u_1 + i \hbar/\tau}  - 
                        \frac{g(x) - 1}{u_2 + u_1 - i \hbar/\tau}   \bigg ) 
  \\   \nonumber
  &-&  \int_{\Delta(T)}^{\infty} \frac{dx}{2} ~  \tanh \bigg ( \frac{x}{2 k_B T} \bigg )
  \\   \label{sigma_chang}
  &&  \times \bigg (  \frac{g(x) + 1}{u_2 - u_1 + i \hbar/\tau}   + 
                      \frac{g(x) - 1}{u_2 + u_1 + i \hbar/\tau}   \bigg ) \, . ~~~ 
\eea
     Here we choose $\tau$ such that $\hbar/ \tau \Delta(0)=\pi\xi_0 /l = 13.61$, corresponding to the
     experimental coherence length $\xi_0 = 39 $ nm and the mean free path $l(T \simeq 9 K) = 9$ nm.
     The normalization constant is $\sigma_L = 1.85 \times 10^{14}\, (\Omega \mbox{m})^{-1}$ corresponding to  
     $\lambda_L(0) = 35 \, \mu$m for niobium \cite{pronin_98}.

\begin{figure}[ht]

\begin{picture}(0,0)(110,663)   

\includegraphics{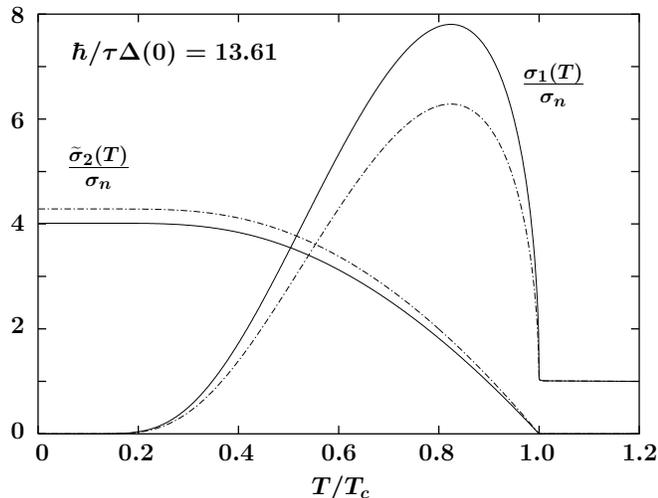}

\end{picture}

\vspace{6.5cm}

\caption{The complex conductivity $\sigma(T) \equiv \sigma_1(T) + i \sigma_2(T)$ as a function of the temperature $T/T_c$
         with $\hbar/\tau \Delta(0)=13.61$ \cite{pronin_98}. 
         The solid line corresponds to \eq{sigma_chang}, and the dashed-dotted line corresponds to \eq{sigma_klein} and
        ${\widetilde{\sigma_2}}(T) \equiv 0.25 \times 10^{-5} \sigma_2 (T)$.
}
\label{sigma_sfa_t_FIG}
\end{figure}

     As the temperature decreases below $T_c$, Cooper pairs will be created. Despite a very small fraction of
     Cooper pair for temperatures just below $T_c$, the imaginary part of the conductivity as given by \eq{sigma_chang}
     exhibits a vast increase (c.f. \fig{sigma_sfa_t_FIG}). 
     Furthermore, due to the modification of the quasi-particle dispersion in the superconducting state, there is
     an increase in $\sigma_1(T)$ as well just below $T_c$. This is the well-known coherence Hebel-Schlichter
     peak \cite{hebel_59}.  In contrast to the simple Gorter-Casimir temperature dependence,
     both Eqs.(\ref{sigma_klein}) and (\ref{sigma_chang}) describe well the presence of the Hebel-Schlichter peak,
     with a peak height less then $8 \, \sigma_n$ for both cases (c.f. \fig{sigma_sfa_t_FIG}),
     at least for the values of the physical parameters under consideration in the present paper.
     In the opposite temperature limit, i.e. $T \ll T_c$, numerical studies of \eq{sigma_chang} show that
     $\sigma_1(T)$ decreases exponentially fast. As seen in \fig{sigma_sfa_t_FIG},  the imaginary part
     of the conductivity, on the other hand, is more or less constant for such temperatures.

\begin{figure}[ht]

\begin{picture}(0,0)(122,665) 

\includegraphics{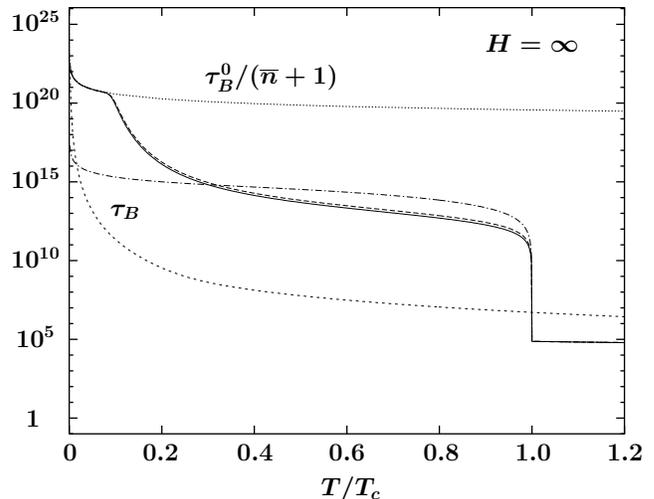}

\end{picture}

\vspace{6.4cm}

\caption{$\tau_B$ of a trapped atom near a superconducting bulk as a function of the temperature $T/T_c$.
         The other relevant parameters are the same as in \fig{tau_sfa_t}.
         The solid line shows the lifetime $\tau_B$ using the microscopic BCS theory, i.e. \eq{sigma_chang}.
         The bold dotted line corresponds to the Mattis-Bardeen theory, i.e. using \eq{sigma_klein}.
         The dashed-dotted line shows the lifetime $\tau_B$ using \eq{sigma_anderson}, with $\hbar/\tau \Delta(0)=13.61$. 
         The dashed line corresponds to a film made of gold described by the dielectric function given by
         \eq{eps_iver}.
         The upper most graph (dotted line) shows to the lifetime $\tau_B^0/( {\bar n} + 1 )$, i.e. the 
         free-space lifetime with $\tau_{B}^0 =1.114\times 10^{25}$.
}
\label{tau_sfa_t_Hinf_FIG}
\end{figure}

     In passing we observe  that there is  only a minor
     difference in $\sigma_2(T)$ as obtained from  Eqs.(\ref{sigma_klein}) and (\ref{sigma_chang}) respectively. 
     For temperatures around the peak value of the Hebel-Schlichter peak, $\sigma_1(T)$ obtained
     from Eq.(\ref{sigma_chang}) is, however, approximatively twenty percent larger than $\sigma_1(T)$ as obtained
     from Eq.(\ref{sigma_klein}). This difference has, nevertheless a small effect on the lifetime $\tau_B$.
     Hence, computing $\tau_B$ using  Eqs.(\ref{sigma_klein}) or (\ref{sigma_chang}) for the complex conductivity,
     we realize that the presence of non-magnetic impurities have no dramatic impact on the lifetime for
     spontaneous emission (see \fig{tau_sfa_t_Hinf_FIG}). A comparison of the values of $\tau_B$ as obtained
     using the two-fluid model for $H=\infty$ as presented in \fig{tau_sfa_t}
     and the corresponding result as shown in \fig{tau_sfa_t_Hinf_FIG} shows, for our set of physical parameters,
     that the two-fluid model overestimates $\tau_B$ with three order of magnitude.

\begin{figure}[ht]

\begin{picture}(0,0)(122,665) 

\includegraphics{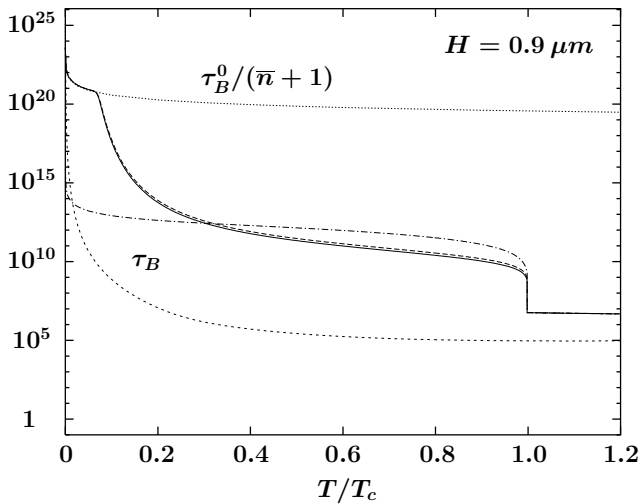}

\end{picture}

\vspace{6.4cm}

\caption{$\tau_B$ of a trapped atom near a superconducting film as a function of the temperature $T/T_c$
         with $H=0.9 \, \mu$m.
         The other relevant parameters and labels are the same as in \fig{tau_sfa_t_Hinf_FIG}.
}
\label{tau_sfa_t_H09_FIG}
\end{figure}

       For finite values of the lifetime $\tau$ and for non-magnetic impurities we can also
       investigate the validity of the two-fluid model approximation  in terms of the lifetime $\tau_B$
       for spontaneous emission processes. As we now will see, there are large deviations between
       the microscopic theory and the two-fluid model approximation, in particular for small
       temperatures. According to Abrikosov and Gor'kov (for an excellent account
       see e.g. Ref.\cite{sadovskii_06} and references cited therein), the density of
       superconducting electrons is given by
\bea   \label{sigma_anderson}
 \frac{n_s(T)}{n_0} \approx \frac{\pi \tau}{\hbar} \, \Delta(T) \, \tanh \left ( \, \frac{\Delta(T)}{2 \, k_B T} \, \right ) ~ ,
\eea 
      provided that $\tau \Delta(0)/\hbar \ll 1$.
      We can now compute the dielectric function \eq{eps_j_2} using \eq{tfm_sigma1_2}.
      We find that $\sigma_2(T)/\sigma_L$ obtained in this way agrees well the
      corresponding quantity obtained from \eq{sigma_chang}. There is, however, a considerable discrepancy
      between the two-fluid expression for $\sigma_1(T)/\sigma_L$ and the corresponding expressions
      obtained from the microscopic theory as given by \eq{sigma_chang}.
      The numerical results for the lifetime in this case are illustrated in the dashed-dotted lines
      in \fig{tau_sfa_t_Hinf_FIG} and in \fig{tau_sfa_t_H09_FIG}.

     Since we are considering low frequencies $0 < \omega \ll \omega_g \equiv 2 \Delta(0)/\hbar$,
     strong coupling effects can now be estimated  by making use of the low-frequency limit of
     the Eliashberg theory \cite{eliashberg_60} and its relation to the BCS theory (see e.g. Ref.\cite{carbotte_90}).
     The so-called mass-renormalization factor $Z_N$, which in general is both frequency and temperature dependent,
     is then replaced by its zero-temperature limit, which for niobium has the value $Z_N \approx 2.1$ \cite{carbotte_90}.
     Using the strong-coupling expressions for the optical conductivity in a suitable form as e.g. given in Ref. \cite{klein_94},
     we then find that the complex conductivity $\sigma (T)/\sigma_n$ is rescaled by
     $\sigma_n \rightarrow \sigma_n/Z_N$ with the lifetime of non-magnetic impurities rescaled by
     $\tau \rightarrow \tau/Z_N$. The change in the lifetime for spontaneous emission can
     then e.g. be inferred from the relation Eq.(\ref{skagerstam06}), and we find only a minor
     decrease of $\tau_B$ by the numerical factor $1/\sqrt{Z_N}\approx 0.69$, which also agrees well with more precise numerical evaluations.

\begin{figure}[ht]

\begin{picture}(0,0)(122,665)   

\includegraphics{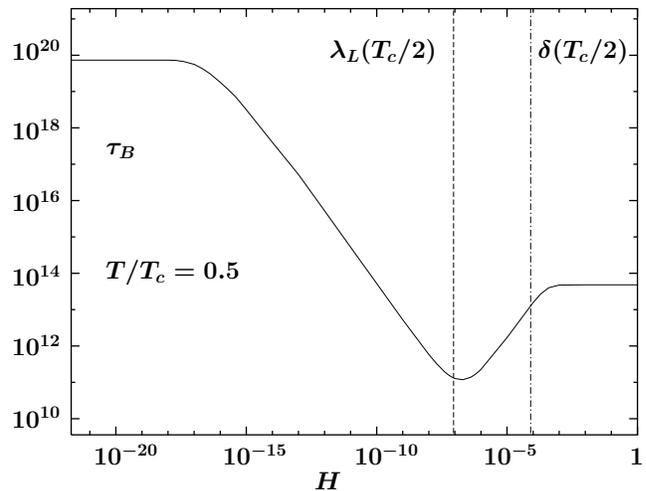}

\end{picture}

\vspace{6.4cm}

\caption{$\tau_B$ as a function of the thickness $H$ of the film. The complex conductivity
         is computed applying \eq{sigma_chang} with $\hbar/\tau \Delta(0) = 13.61$.
         Other relevant parameters are the same as in the solid line in \fig{tau_sfa_t}.
         In the limit $H=0$, i.e. no slab at all, the lifetime is simply $\tau_B^0/( n + 1 )
         = 7.34 \cdot 10^{19}$ s for the parameters under consideration.
         The skin depth $\delta(T_c/2) = 80.5 \, \mu$m is indicated by a dashed-dotted line,
         and the London length $\lambda_L(T_c/2) = 89.2 \,$nm is indicated by a dashed line.
}
\label{tau_sfa_H_FIG}
\end{figure}

    The lifetime for spontaneous emission exhibits a minimum with respect to variation of the thickness $H$
    of the superconducting film. This fact is illustrated in \fig{tau_sfa_H_FIG}.    
    Below the minimum at $H_{min} \approx 0.1 \, \mu$m, 
    a decrease of the thickness $H$ leads to an increase of lifetime in proportion to $H^{-1}$. This happens despite the
    growth in polarization noise because the region generating the noise is becoming thinner
    as it is limited by $H$, and not $\lambda_L(T)$.
    Eventually, the lifetime reaches the free-space lifetime is $\tau^0_B$ as $H$ tends to zero.
    On the other hand, for large $H$, i.e. $H \gg \delta(T)$, the lifetime is constant with respect to $H$,
    giving the same result as for an infinite thick slab.
    In the region between, i.e. $\lambda_L(T) \lesssim H \lesssim \delta(T)$, the lifetime is proportional to $H$. 
    Numerical studies show that a non-zero $\sigma_2(T)$ is important for 
    a well pronounced minimum of $\tau_B$ as a function of $H$.

     Some experimental work has been done using a superconducting body, e.g. Nirrengarten {\it et al.} \cite{haroche_06}. 
     Here cold atoms were trapped near a superconducting surface. At the distance of $440 \, \mu$m from the chip surface,
     the trap lifetime reaches $115$ s at low atomic densities and with a temperature $40 \, \mu$K of the chip. 
     We believe the vast discrepancy between this experimental value and our theoretical calculations must rely on
     effects that we have not taken into account in our analysis. 
     The use of a thin superconducting film may lead to the presence vortex motion 
     and pinning effects in (see e.g. Refs.\cite{larkin_94,villegas_05}). The presence of vortices will
     in general modify the dielectric properties of the dielectric body. If we, as an example,
     consider a vortex system in the liquid phase in a finite slab geometry, one expects a strongly
     temperature dependent $\sigma_1 (T)$ with a peak value $\sigma_1 (T) \simeq 1.3\times 10^7/H^2[\mu m]\nu [kHz] \Omega m$ \cite{larkin_94}.
     Close to this peak $\sigma_1 (T) \simeq \sigma_2 (T)$, and for $\nu \simeq 560 \, kHz$ we find a lifetime
     for spontaneous emission two orders of magnitude larger than a film made out of gold with the same geometry.
     It is an interesting possibility that spontaneous emission processes  close to thin superconducting films
     could be used for an experimental study of the physics of vortex condensation. This possibility has also
     been noticed in a related consideration, which has appeared during the preparation of the
     present work \cite{fermani_2007}. There are also fabrication issues concerning the Nb-O
     chemistry \cite{halbritter_99} which may have an influence on the lifetime for spontaneous emission.

     To summarize, we have studied the rate for spontaneous photon emission, due to a magnetic
     spin-flip transition, of a two-level atom in the vicinity of a normal conducting metal or a superconductor. 
     Our results confirms the conclusion in Ref.\cite{rekdal_06},
     namely that the corresponding magnetic spin-flip lifetime will be boosted by several orders of magnitude by
     replacing a normal conducting film with a superconducting body. This conclusion holds when
     describing the electromagnetic properties of the superconductivity in terms of a simple two-fluid model
     as well as in terms of a more detailed  and precise microscopic Mattis-Bardeen and
     Abrikosov-Gor'kov-Khalatnikov theory. 
     For the set of physical parameters as used in Ref.\cite{rekdal_06} it so happens, more or less by chance,
     that the two-fluid model results agree well with the results from the microscopic BCS theory.
     We have, however, seen that even though the two-fluid model gives a qualitatively correct physical picture
     for spontaneous photon emission, it,  nevertheless, leads to large quantitative  deviations when
     compared to a detailed  microscopic treatment. We therefore have to resort to the microscopic
     Mattis-Bardeen \cite{mattis_58} and Abrikosov-Gor'kov-Khalatnikov \cite{gorkov58} theory in order
     to obtain precise predictions. We have also show that non-magnetic impurities as well as strong-coupling
     effects have no dramatic impact on the rate for spontaneous photon emission. Vortex condensation in thin
     superconducting films may, however, be of great importance. Finally, we stress the close relation
     between the spin-flip rate for spontaneous emission and the complex conductivity, which indicates
     a new method to experimentally study the electrodynamical properties of a superconductor or a normal conducting metal.
     In such a context the parameter dependence for a bulk material as given by Eq.(\ref{skagerstam06}) may be useful.

\begin{center}ACKNOWLEDGEMENTS
\end{center}

     This work has been supported in part by the Norwegian University of Science and Technology (NTNU)
     and the Norwegian Research Council (NFR). One of the authors (B.-S.S.) wishes to thank Professors Y. Galperin,
     T.H. Johansen, V. Shumeiko and G. Wendin for fruitful discussions.


\end{document}